# Fe-C and Fe-H systems at pressures of the Earth's inner core.

Zulfiya G. Bazhanova, Artem R. Oganov, Omar Gianola

## Abstract

The solid inner core of the Earth is predominantly composed of iron alloyed with several percent Ni and some lighter elements, Si, S, O, H, and C being the prime candidates. To establish the chemical composition of the inner core it is necessary to find the range of compositions that can explain its observed characteristics. Recently, there have been a growing number of papers investigating C and H as possible light elements in the core, but the results are contradictory. Here, using *ab initio* simulations, we study the Fe-C and Fe-H systems at inner core pressures (330-364 GPa). Using the evolutionary structure prediction algorithm USPEX, we have determined the lowest-enthalpy structures of possible carbides (FeC, $Fe_2C$, $Fe_3C$, $Fe_4C$, $FeC_2$, $FeC_3$, $Fe_2C_4$ and $Fe_7C_3$) and hydrides ($Fe_4H$, $Fe_3H$, $Fe_2H$, FeH, $FeH_2$, $FeH_3$, $FeH_4$) and have found that $Fe_2C$ (space group *Pnma*) is the most stable iron carbide at pressures of the inner core, while FeH, $FeH_3$ and $FeH_4$ are stable iron hydrides at these conditions. For $Fe_3C$, the cementite structure (space group *Pnma*) and the *Cmcm* structure recently found by random sampling are less stable than the *I-4* and *C2/m* structures found here. We have found that $FeH_3$ and $FeH_4$ adopt chemically interesting thermodynamically stable structures, in both compounds containing trivalent iron. We find that the density of the inner core can be matched with a reasonable concentration of carbon, 11-15 mol % (2.6-3.7 wt. %) at relevant pressures and temperatures. This concentration matches that in CI carbonaceous chondrites and corresponds to the average atomic mass in the range 49.3-51.0, in close agreement with inferences from the Birch's law for the inner core. Similarly made estimates for the maximum hydrogen content are unrealistically high, 17-22 mol.% (0.4-0.5 wt. %), which corresponds to the average atomic mass in the range 43.8-46.5. We conclude that carbon is a better candidate light alloying element than hydrogen.

## 1. Introduction

The problem of determination of the chemical composition of the Earth's core has fascinated geoscientists, physicists, chemists and materials scientists for several decades. It is clear that iron-based alloys are the dominant components of the Earth's solid inner core and liquid outer core, but according to seismic models, the density of the core is several percent lower than the density of pure iron or an iron-nickel alloy at relevant pressures and temperatures [1-3]. From the Birch's law [4] one can deduce that the mean atomic mass in the core is ~49 [5] – for pure iron it is 55.85. To explain these differences, one has to allow for ~10-20 mol% of a lighter element [1-3, 6, 7]. Poirier [8], considering four postulates of Stevenson [2], has specified Si, S, O, H, and C as the likeliest candidates. Initially, relatively few papers considered carbon and hydrogen as major light elements in the core, but recently the number of studies has increased dramatically. There is little consistency between published works on this issue, and here we want to analyze it avoiding assumptions and extrapolations that have previously led to long-standing controversies.

According to Wood [9], "carbon is extremely abundant in the Solar System (10 × Si, 20 × S) in C1 carbonaceous chondrites (3.2 wt%) and it dissolves readily in liquid Fe at low pressures (4.3 wt% at 1420 K). Despite these properties it is rarely considered a potential light element in the Fe-rich core, because it is volatile, even at low temperatures as CO". Wood [9] concluded that carbon should be considered as a light component of the core, because its volatility decreases and solubility in liquid iron increases under pressure. However, Poirier [8] noted that even at high pressures carbon's solubility in iron will remain insufficient to explain the density deficit in the core. Tingle [10] proposed that carbon has been incorporated in the core during its formation, and as supporting evidence used the observed large amounts of carbon in iron meteorites, as well as experiments on its high-pressure solubility in liquid iron [9, 11]. Estimates of carbon content in the inner core range from 0.2 wt.% [12] to 4 wt.% [13]. Anisichkin [14], based on shock-wave and seismic data, advocated carbon as a major light element (as much as 10 wt%) in the core, partly in the diamond phase. However, from first-principles calculations [15, 16] it is clear that diamond cannot be present in the core, because diamond reacts with iron at high pressures, forming iron carbides. Experimental results of Tateno et al. [17], obtained at pressures and temperatures of the Earth's inner core, indicate low solubility of carbon in iron and coexistence of iron with some iron carbide at these conditions. The concentration of carbon in the inner core has been evaluated using the equations of state of these carbides. Li et al. [18] studied compressibility of cementite at pressures up to 30.5 GPa, and Scott et al. [19], using the experimental equation of state of cementite measured up to 73 GPa at 300 K, have concluded that carbon can be a major light alloying element in the core. Using *ab initio* simulations, Vocadlo et al. [20] found that collapse of magnetism at ~60 GPa has a major effect on the equation of state of cementite. The magnetic collapse was experimentally observed [21] at ~45 GPa; the observation of a large increase of incompressibility lead to the conclusion that the presence of carbon will not improve the match with the observed seismic properties of the inner core. Based on their analysis of the thermal equation of state of $Fe_3C$ and Fe, Huang et al. [15] concluded that carbon cannot be a major element of the core. The same conclusion was reached by Sata et al. [22], whereas Nakajima et al. [23], Fiquet et al. [24] and Gao et al. [25] have arrived at the opposite conclusion, i.e. that carbon can be a the major light alloying element in the inner core. Curiously, the above mentioned discrepant conclusions were reached based on the equation of state of the same phase, cementite, which until recently was assumed to be stable at conditions of the inner core.

Yes, this assumption deserves to be challenged as it is based essentially on nothing and turned out to be incorrect. Related to this issue, the experiments [26] at 2200-3400 K and 25-70 GPa witnessed stability of $Fe_3C$ at those conditions. However, on the basis of their experimental phase diagrams, Lord et al. [27] demonstrated that most likely cementite is irrelevant for the inner core and one should consider $Fe_7C_3$ instead. Nakajima et al.[28] studied this phase and its equation of state at pressures up to 71.5 GPa, while Mookherjee et al. [29] computed it up to the pressures of the inner core using *ab initio* simulations. Both works concluded that the incorporation of carbon does provide a good match for the density of the inner core. Mookherjee et al. [29] estimated the amount of carbon needed to for this match to be 1.5 wt.% (6.6 mol.%). But what if at higher pressures, corresponding to the actual conditions of the inner core, yet another composition or structure is stable? Weerasinghe et al. [30] used random sampling approach [31] and suggested that $Fe_2C$ is more stable than other compositions, but as they admitted that random sampling was not powerful enough to predict structure of $Fe_7C_3$, a re-examination of the Fe-C system is warranted. We investigate this question here and give new estimates of the maximum carbon content in the inner core.

Hydrogen is the element with the highest abundance in the solar system ($10^4$xSi). Therefore, it seems logical to suppose that hydrogen could be the main element responsible for the density deficit observed in the Earth's core. Estimates of Stevenson [2] showed that the presence of FeH could explain the density of the inner core, but low solubility of hydrogen in iron at atmospheric pressures makes this possibility less likely. Further studies pointed out that at high pressure the solubility of hydrogen in iron increases considerably [32] and the iron hydride phase $FeH_x$ (with a stoichiometry approaching 1:1 and a double hexagonal close packed (dhcp) structure, with interstitial hydrogen in octahedral sites) could be stable [33]. Depending on experimental conditions, different close-packed iron hydride phases were synthesized – dhcp, hcp, and fcc [34, 35, 36], and at least up to 80 GPa the most stable phase has the dhcp structure [34, 37]. Skorodumova et al. [38] demonstrated by *ab initio* calculations that, at high pressures, hydrogen stabilizes close-packed structures (hcp, dhcp and fcc) of iron and fills the octahedral voids in the structure. Isaev et al. [39] predicted that the fcc phase of FeH will be stable above 83 GPa.

X-ray diffraction experiments performed in the diamond anvil cell at room temperature and pressures up to 80 GPa [37] showed two discontinuities (at 30 and 50 GPa) in the c/a axial ratio of $FeH_x$ and inferences were made to the lower hydrogen content (0.12-0.48 wt.%) than previously [40, 33] thought. Synchrotron Mössbauer measurements of [41] demonstrated the loss of magnetism at 22 GPa, i.e. well far below the ab initio determined ferromagnetic-nonmagnetic transition at ~60 GPa [42].

Iron hydrides could be formed in the Earth's core by the reaction of iron with water during the early stages of the Earth's accretion [43, 40, 33]: $(2+x)Fe+H_2O=2FeH+Fe_xO$. Yagi and Hishinuma [44] studied the interaction between hydrogen and iron in the $Fe-Mg(OH)_2-SiO_2$ system in the pressure range 2.2-4.9 GPa and temperatures up to 1350 °C, where the water was supplied by the decomposition of $Mg(OH)_2$ brucite. They observed that at 2.2 GPa and above 550 °C, iron hydride (however with a chemical composition estimated to be $FeH_{0.3}-FeH_{0.4}$) was formed. Therefore, the formation of iron hydride at pressures as low as 2.2 GPa implies that if water was present in the proto-Earth together with silicates and iron, iron hydride would be formed at relatively shallow depths. Assuming a primordial Earth characterized by a hydrous magma ocean, Okuchi [45] calculated that, if the pressure at the bottom of the magma ocean was higher than 7.5 GPa, then more than 95 mol% of the water could react with iron to form $FeH_x$, which later sank to build the proto-core. Therefore, it seems that currently there are no compelling geochemical reasons against the presence of significant amounts of either carbon or hydrogen in the Earth's inner core.

Here we consider this problem from the point of view of mineral physics. It has been suggested [46] that hydrogen and carbon cannot be simultaneously present in the core. With this in mind, we consider separately the Fe-C and Fe-H systems, finding the stable iron carbides and hydrides at pressures of the inner core. We consider crystal chemistry of the stable iron carbides and hydrides and determine, on the basis of the most accurate available data, how much carbon or hydrogen is needed in order to match the observed density of the inner core of the Earth. Our calculations are based on the evolutionary crystal structure prediction method USPEX [47-50] and density functional theory [51, 52] within the generalized gradient approximation (GGA) [52]. These calculations successfully reproduce the known facts about these systems and predict hitherto unknown crucial pieces of information about the behavior of carbon and hydrogen in the Earth's inner core.

## 2. Methodology

As a simple test of the performance of the GGA, Fig. 1 compares theoretical (at 0 K) and experimental (at 300 K) equations of state of $Fe_3C$, $Fe_7C_3$ and FeH. Calculations performed here are performed only for non-magnetic states, because all experimental and theoretical evidence indicates collapse of magnetism in the Fe-C and Fe-H systems at several tens of Gigapascals – well below the pressures of the inner core (330-364 GPa). From Fig. 1 one can see that agreement between theoretical and experimental equations of state is quite good, especially at pressures above ~100 GPa, and improves on increasing pressure.

Using the USPEX method [47-50] combined with GGA calculations, we found the lowest-enthalpy crystal structures at pressures of 300 GPa and 400 GPa corresponding to compositions FeC, $Fe_2C$, $Fe_3C$, $Fe_4C$, $FeC_2$, $FeC_3$, $FeC_4$ and $Fe_7C_3$ for the Fe-C system, and $Fe_4H$, $Fe_3H$, $Fe_2H$, FeH, $FeH_2$, $FeH_3$, $FeH_4$ for the Fe-H system. Such calculations for pure Fe and C [47] and for H (Oganov, unpublished) have produced the known lowest energy structures – hcp-Fe, diamond and *Cmca*-12-H (at 300 K) and *Cmca*-H (at 400 GPa), respectively, in agreement with available experimental (e.g. for iron [54, 55, 17]) and theoretical [56] evidence. At each composition, for a given pressure, we identified the most favorable crystal structure and computed its enthalpy of formation from the elements. These enthalpies, normalized per atom, are given in Fig. 2, where stable compositions form a convex hull (i.e. a set of points lying below all lines joining any pair of other points on the diagram).

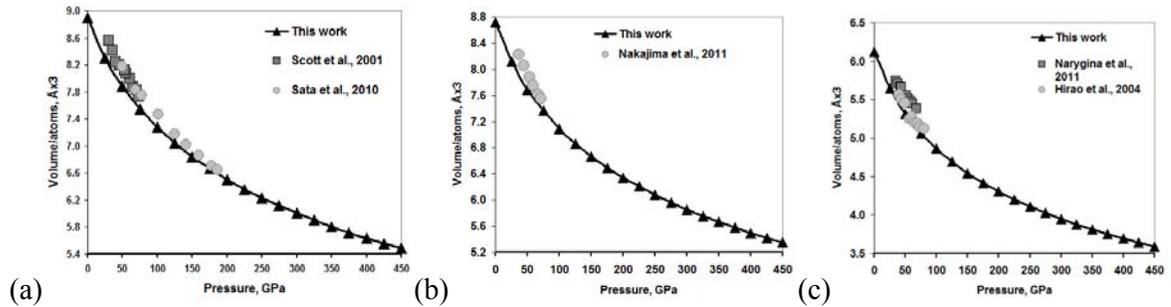

(a)   (b)   (c)

**Fig. 1. Comparison of the theoretical (at 0 K) and experimental (at 300 K) equations of state:** (a) $Fe_3C$ cementite, in comparison with high-pressure data of Scott et al. [19] and Sata et al. [22]; (b) $P6_3mc$-$Fe_7C_3$, in comparison with high-pressure data of Nakajima et al. [23]; (c) rocksalt-type FeH, in comparison with high-pressure data of Narygina et al. [46] for the same phase, and of Hirao et al. [37] for the related dhcp-FeH phase. Theoretical data are for the non-magnetic state, which has been shown to be stable above 67 GPa for $Fe_7C_3$ [29] and above 60 GPa [20] or 25 GPa [21] for $Fe_3C$, for FeH at 22 GPa [**Error! Reference source not found.**] or 60 GPa [**Error! Reference source not found.**].

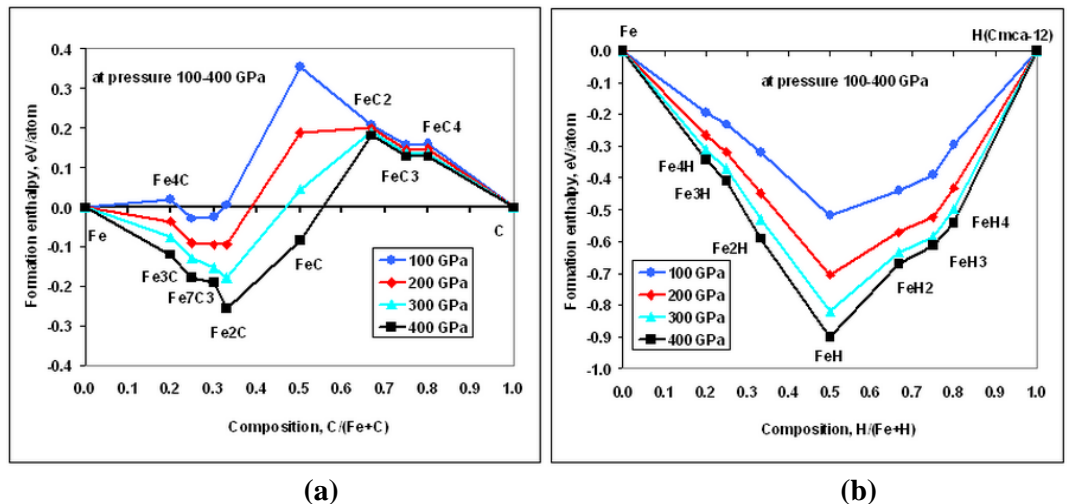

(a)   (b)

**Fig. 2. Predicted enthalpies of formation of (a) Fe-C and (b) Fe-H compounds.**

Structure prediction calculations were performed for $Fe_4C$ (and $Fe_4H$) with 10 and 15 atoms/cell, for $Fe_3C$ (and $Fe_3H$) with 12 and 16 atoms/cell, for $Fe_7C_3$ with 10 and 20 atoms/cell, for $Fe_2C$ (and $Fe_2H$) with 9 and 12 atoms/cell, for FeC (and FeH) with 12 and 16 atoms/cell, $FeC_2$ (and $FeH_2$) with 9 and 12 atoms/cell, $FeC_3$ (and $FeH_3$) with 12 and 16 atoms/cell, and $FeC_4$ (and $FeH_4$) with 10 and 15 atoms/cell. A typical USPEX simulation included 30-40 structures per generation, the lowest-enthalpy 60% of which were used for producing the next generation of structures (70% of the offspring produced by heredity, 10% by permutation, and 20% by lattice mutation). All structures produced by USPEX were relaxed and their enthalpy computed using density functional theory within the generalized gradient approximation (GGA) [52] and employing the projector-augmented wave method [57,58], as implemented in the VASP code [59]. We used PAW potentials with an [Ar] core (radius 2.3 a.u.) and [He] core (radius 1.52 a.u.) for Fe and C atoms, respectively, and a PAW potential for H with core radius of 1.1 a.u.. A plane-wave kinetic energy cut-off of 600 eV and 350 eV was used for Fe-C and Fe-H systems, respectively, and demonstrated to give excellent convergence of stress tensors and structural energy differences. During structure relaxations done within USPEX simulations, we used homogeneous Monkhorst-Pack k-point meshes with reciprocal-space resolution of $2\pi x 0.08$ Å$^{-1}$ and Methfessel-Paxton electronic smearing [60] with σ=0.1 eV. Having identified several lowest-enthalpy structures using USPEX, we recalculated their enthalpies in a range of pressures using a denser sampling of the Brillouin zone with the resolution of $2\pi x 0.05$ A$^{-1}$. In addition to structures found by USPEX, we also considered some experimentally known structures – *Pnnm* for Fe$_2$C, cementite (*Pnma*) and bainite (*P*6$_3$22) for Fe$_3$C (cementite was also seen in USPEX simulations), *P*-43*m* for Fe$_4$C, and Cr$_7$C$_3$-type (type D101, space group *Pnma*) and *P*6$_3$*mc* for Fe$_7$C$_3$ (which was also found in USPEX calculations). In all cases USPEX successfully produced structures that have the lowest enthalpy among all known or hypothetical structures. This gives us confidence in the reliability of the results discussed below.

## 3. Structures and compositions of stable iron carbides at ultrahigh pressures.

For almost all Fe-C alloy compositions, at inner core pressures we find more stable structures than those known experimentally at lower pressures. The new structures are shown in Fig. 3. For Fe$_3$C, experiments [22] show cementite to remain stable at least to 187 GPa (the highest pressure probed in the experiments [22]), which has led researchers to believe the cementite structure (space group *Pnma*, see Table 1) to be stable all the way up to the pressures of the inner core. Here we establish the upper limit of its stability as 310 GPa. Above 310 GPa (and this includes the actual pressures of the Earth's inner core, 330-364 GPa) USPEX found two more stable structures – with space groups *I*-4 (stable at 310-410 GPa) and *C2/m* (stable above 410 GPa) (Fig. 4). One of the phases experimentally known at 1 atm, bainite (space group P6$_3$22), is less stable than cementite, *I*-4 or *C2/m* phases of Fe$_3$C at all pressures. The *Cmcm* structure, predicted using random sampling as the most stable structure of Fe$_3$C above 326 GPa [30], turns out to be less stable than the structures predicted here at all pressures. At 300 GPa it is 18 meV/f.u less stable than cementite and 7 meV/f.u. than the *I*-4 structure. At 350 GPa, it is 16 meV/f.u. higher in enthalpy than the *I*-4 phase. At 400 GPa it is 14 meV/f.u. less stable than the *C2/m* structure, and by 19 meV/f.u. than the *I*-4 phase. This failure of the random sampling approach is well known and has been documented for several other systems, such as SiH$_4$, SnH$_4$, N, for which evolutionary simulations using USPEX found more stable structures [50]. Since both for Fe$_7$C$_3$ and Fe$_3$C the random sampling calculations of Weerasinghe et al. [30] failed to find stable structures, their other conclusions on Fe-C phases are in doubt too.

For Fe$_2$C, the *Pnnm* structure, experimentally known at 1 atm, at core pressures turns out to be much less stable than the *Pnma* structure predicted by USPEX (Table 1) – by 0.59 eV/atom at 300 GPa and 0.75 eV/atom at 400 GPa. For Fe$_4$C, the *P*2$_1$/*m* and *I*4/*m* structures (Table 1) predicted by USPEX at 300 GPa and 400 GPa, respectively, are vastly (by 1.5-1.7 eV/atom) superior to the experimentally known *P*-43*m* structure with 5 atoms in the unit cell. The stability of the *Pnma* phase of Fe$_2$C is consistent with the results of Weerasinghe et al. [30].

Fe$_7$C$_3$ is unique among the compounds considered here in that the structure known experimentally at 1 atm (space group *P*6$_3$*mc*, Table 1) is also stable at pressures of the Earth's inner core. This structure was successfully found in our USPEX simulations, in contrast to the previous random sampling calculations [30]. This structure is more stable than the other phase of Fe$_7$C$_3$ experimentally known at 1 atm, with the *Pnma* space group and 40 atoms/cell.

For the other compositions (FeC, FeC$_2$, FeC$_3$, FeC$_4$) we are not aware of any experimentally known phases. None of these compositions are found to be stable with respect to decomposition into the elements or into a mixture of C and Fe$_2$C in the investigated pressure range. In some cases, e.g. FeC$_4$, we observed phase separation in the predicted lowest-enthalpy structure into layers of iron and diamond within one simulation cell.

To summarize, using USPEX we have found new lowest-enthalpy structures for Fe$_4$C (*P*2$_1$/*m* and *I*4/*m*), Fe$_3$C (*I*-4 and *C2/m*) and Fe$_2$C (*Pnma*) at inner core pressures (Fig. 3). These structures are extremely interesting; in these, carbon atoms have 8- and 9-fold coordination. The Fe$_4$C-*P*2$_1$/*m* structure shows Fe-C layering and can be described as hexagonal close packing of iron atoms strongly distorted by the insertion of carbon atoms, while Fe$_4$C-*I*4/*m* structure can be obtained from a body-centered cubic structure. Some compositional layering can be seen in the Fe₃C-*C2/m* structure, while the Fe₃C-*I*-4 structure contains remarkable tetragonal channels. The Fe₂C-*Pnma* structure, which we predict to be the stable iron carbide at pressures of the inner core, displays no compositional layering and contains carbon atoms in the 8-fold coordination.

From our data, it is clear that Fe₃C is not a stable carbide at pressures of the Earth's inner core – contrary to the common belief. Fig. 2 (a) shows the enthalpies of formation of all studied Fe-C compounds. Using the convex hull construction, it is easy to see from this graph that among different iron carbides:

at 100 GPa, Fe₃C and Fe₇C₃ are stable

at 200 GPa, Fe₃C and Fe₂C are stable

at 300 GPa and 400 GPa, only Fe₂C is stable

These results clearly indicate that the traditional thinking, based on Fe₃C or Fe₇C₃ as the most stable iron carbide at Earth's inner core pressures, must be abandoned.

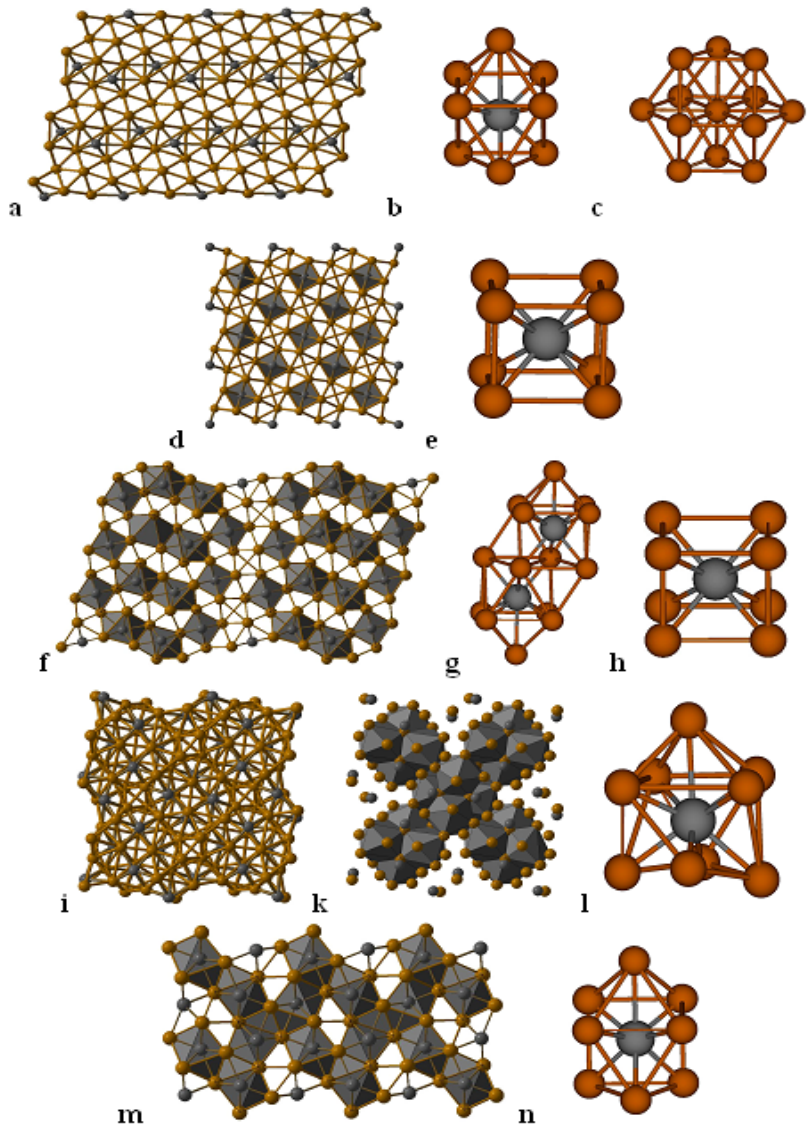

**Fig. 3. New structures of iron carbides found in this work.** Brown circles correspond to iron atoms, dark grey circles to carbon atoms. (a) Fe₄C (*P2₁/m*) structure and (b) 8-coordinate environment of the carbon and (c) 12-coordinate (hexagonal cuboctahedral) environment of the iron atoms in it; (d) Fe₄C (*I4/m*) structure and (e) 8-coordinate environment of the carbon atom in it; (f) Fe₃C (*C2/m*) structure and (g,h) 9- and 8-coordinate environments of carbon atoms in it;

(i-k) ball-and-stick and polyhedral representations of the Fe$_3$C (*I*-4) structure and (l) 9-coordinate environment of carbon atoms in it, (m) Fe$_2$C (*Pnma*) structure and (n) 8-coordinate environment of a carbon atom in it.

**Table 1. Structural parameters of some of the phases found by USPEX.**

| | Wyckoff position | x | y | z |
|---|---|---|---|---|
| **Fe$_2$C (space group *Pnma*) at 300 GPa.** | | | | |
| a=5.169 Å, b=2.232 Å, c=5.945 Å | | | | |
| Fe | 4c | 0.834 | 0.25 | 0.951 |
| Fe | 4c | 0.499 | 0.75 | 0.839 |
| C | 4c | 0.796 | 0.75 | 0.176 |
| **Fe$_7$C$_3$ (space group *P6$_3$mc*) at 300 GPa.** | | | | |
| a=b=5.987 Å, c=3.773 Å | | | | |
| Fe | 2b | 1/3 | 2/3 | 0.253 |
| Fe | 6c | 0.460 | 0.540 | 0.722 |
| Fe | 6c | 0.122 | 0.878 | 0.428 |
| C | 6c | 0.191 | 0.809 | 0.00 |
| **Fe$_3$C cementite (space group *Pnma*) at 300 GPa.** | | | | |
| a=4.325 Å, b=5.778 Å, c=3.843 Å | | | | |
| Fe | 4c | 0.022 | 0.75 | 0.368 |
| Fe | 8d | 0.191 | 0.558 | 0.843 |
| C | 4c | 0.885 | 0.75 | 0.942 |
| **Fe$_3$C (space group *C2/m*) at 400 GPa.** | | | | |
| a=7.321 Å, b=2.155 Å, c=11.720 Å, β=104.76° | | | | |
| Fe | 4i | 0.485 | 0.00 | 0.778 |
| Fe | 4i | 0.731 | 0.00 | 0.722 |
| Fe | 4i | 0.644 | 0.00 | 0.529 |
| Fe | 4i | 0.495 | 0.50 | 0.626 |
| Fe | 4i | 0.380 | 0.50 | 0.899 |
| Fe | 4i | 0.131 | 0.50 | 0.953 |
| C | 4i | 0.682 | 0.50 | 0.839 |
| C | 4i | 0.304 | 0.00 | 0.622 |
| **Fe$_3$C (space group *I*-4) at 400 GPa.** | | | | |
| a=b=7.104 Å, c=3.555 Å | | | | |
| Fe | 8g | 0.357 | 0.481 | 0.738 |
| Fe | 8g | 0.186 | 0.285 | 0.487 |
| Fe | 8g | 0.412 | 0.104 | 0.489 |
| C | 8g | 0.527 | 0.292 | 0.760 |
| **Fe$_4$C (space group *P2$_1$/m*) at 400 GPa.** | | | | |
| a=5.293 Å, b=2.196 Å, c=5.423 Å, β=103.01° | | | | |
| Fe | 2e | 0.876 | 0.75 | 0.844 |
| Fe | 2e | 0.184 | 0.75 | 0.511 |
| Fe | 2e | 0.671 | 0.25 | 0.074 |
| Fe | 2e | 0.522 | 0.25 | 0.678 |
| C | 2e | 0.102 | 0.25 | 0.743 |
| **Fe$_4$C (space group *I4/m*) at 400 GPa.** | | | | |
| a=b=5.188 Å, c=2.132 Å | | | | |
| Fe | 8h | 0.423 | 0.772 | 0.00 |
| C | 2a | 0.00 | 0.00 | 0.00 |
| **FeH (space group _Fm-3m_) at 300 GPa.** a=3.162 Å | | | | |
| Fe | 4a | 0.00 | 0.00 | 0.00 |
| H | 4b | 0.50 | 0.50 | 0.50 |
| **FeH₃ (space group _Pm-3m_) at 300 GPa.** a=2.215 Å | | | | |
| Fe | 1a | 0.00 | 0.00 | 0.00 |
| H | 3c | 0.00 | 0.50 | 0.50 |
| **FeH₃ (space group _Pm-3n_) at 400 GPa.** a=2.702 Å | | | | |
| Fe | 2a | 0.00 | 0.00 | 0.00 |
| H | 6d | 0.25 | 0.50 | 0.00 |
| **FeH₄ (space group _P2₁/m_) at 300 GPa.** a=3.479 Å, b=3.062 Å, c=2.331 Å, β=101.63° | | | | |
| Fe | 2e | 0.252 | 0.25 | 0.551 |
| H | 4f | 0.370 | 0.958 | 0.118 |
| H | 2e | 0.198 | 0.75 | 0.561 |
| H | 2a | 0.00 | 0.00 | 0.00 |

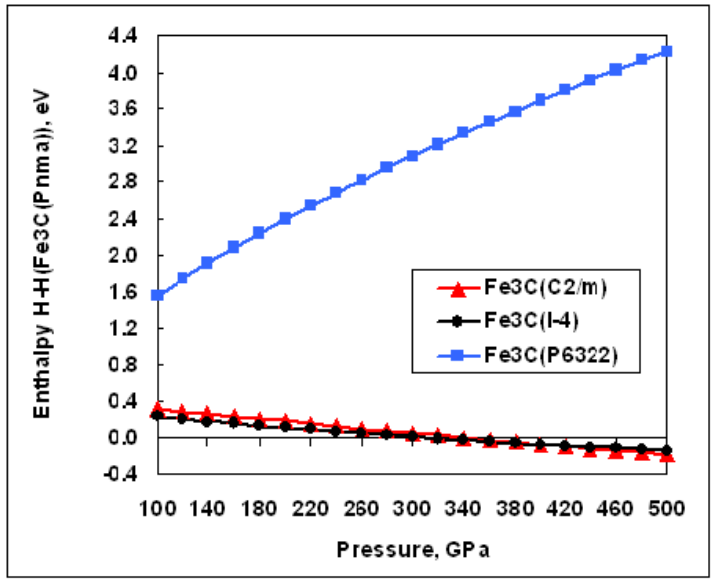

**Fig. 4. Enthalpies of Fe₃C polymorphs - _C2/m_, _I_-4 and _P6₃22_ as a function of pressure.** Enthalpies are shown per formula unit and relative to cementite (_Pnma_).

## 4. Structures and compositions of stable iron hydrides at ultrahigh pressures.

In agreement with previous works (e.g. [33]) we found that FeH is stable at pressures of the inner core, but in addition to it phases with compositions FeH₃ and FeH₄ are also stable when hydrogen fugacity is high, and phases with compositions Fe$_X$H (x>1), such as Fe₄H, are only marginally less stable than the isochemical mixture of Fe and FeH. Crystal chemistry for Fe/H>1 and <1 is very different, and below we consider both, although for the inner core only Fe/H>1 is relevant.

For Fe$_X$H with x≥1, the lowest-enthalpy structures have Fe atoms forming close-packed sublattices (fcc in FeH and hcp in Fe₄H), with H atoms filling the octahedral voids. The stable structure of FeH at pressures of the inner core is of rocksalt type, with the fcc-packing of iron atoms, in agreement with previous calculations [38]. One could expect that hydrogen will stabilize close packed structures of iron in the Earth's core.

The predicted metastable structures allow us to draw some conclusions on the energetics of H-H interactions and effects of partial hydrogen incorporation into the structure of iron. For illustration, let us consider Fe₄H, the lowest-enthalpy structure of which has space group $P$-$3m1$ and can be described as an hcp packing of iron atoms with H atoms filling one in four octahedral layers (Fig. 5a). There are very many alternative and slightly less stable ways of occupying one quarter of the octahedral voids in close-packed structures, and Fig. 5 illustrates some of them. For this composition, the most stable structures are based on the hcp-packing of Fe atoms (other packings have enthalpies >20 meV/atom higher). Within the hcp-packing of Fe atoms, H atoms show the tendency to segregate into layers, which implies that incorporation of moderate amounts of hydrogen into iron will increase the anisotropy of the hcp structure. Face sharing of the H-centered octahedral tends to be avoided – structures with the maximum extent of face sharing (Fig. 5c) are destabilized by ~20 meV/atom, whereas structures fully avoiding face sharings (such as the one shown in Fig. 5b) have enthalpies within 8 meV/atom of the ground state.

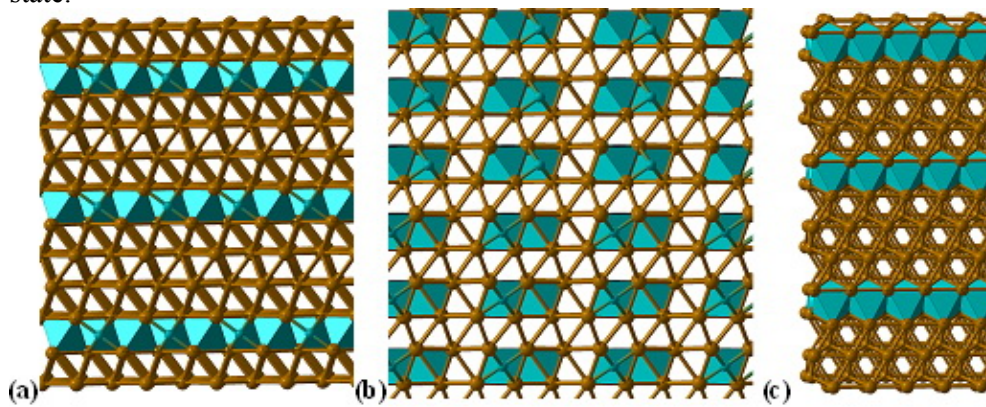

**Fig. 5. Structures of Fe₄H:** (a) lowest-enthalpy $P$-$3m1$ structure and (b,c) higher-energy structures of Fe₄H based on different arrangements of the H atoms in the hcp structure. Turquoise octahedra highlight positions of the H atoms, whereas iron atoms are shown by the orange spheres.

For Fe$_X$H with x<1, FeH₃ and FeH₄ are thermodynamically stable at the pressures of the inner core, are of substantial chemical interest and are built on very different principles. The stable structure of FeH₃ at 300 GPa belongs to the Cu₃Au structure type (space group $Pm3m$, Fig. 6b), while the structure preferred at 400 GPa is of Cr₃Si type (also known as A15 type, space group $Pm3n$, Fig. 6c). The Cu₃Au-type structure is a superstructure of the fcc type, with all atoms in the 12-fold coordination, suggesting that in this structure of FeH₃ iron and hydrogen atoms have comparable sizes (unlike in FeH and Fe$_X$H compounds with x>1, where hydrogens were small enough to fit the octahedral voids of the close-packed iron sublattice). In the Cr₃Si-type structure, known for many superconductors (such as Nb₃Sn and Nb₃Ge, which held record-high superconducting Tc values before the advent of cuprate superconductors), coordination numbers are more ambiguous, but are at any rate similar for Fe and H atoms; the most interesting feature is the presence of FeH₁₂ icosahedral units. The closest H-H distance at 400 GPa is 1.35 Å.

Stability of FeH₄ above ~180 GPa is certainly surprising and raises the question whether Fe in this compound is in the unusual tetravalent state. The stable structure of FeH₄ found by USPEX calculations at 300 GPa and 400 GPa (Fig. 6d) is complex, has low symmetry (space group $P2_1/m$) and belongs to a new structure type. It contains many H-H bonds, the shortest of which (at 400 GPa) are 1.16 Å long.

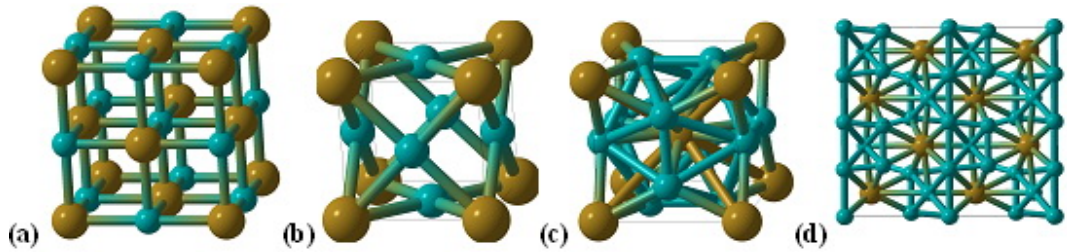

**Fig. 6. Structures of stable high-pressure iron hydrides: (a) rocksalt-type FeH, (b) *Pm3m* and (c) *Pm3n* phases of FeH₃ and (d) *P2₁/m* phase of FeH₄.**

To resolve the valence state of Fe atoms in FeH₄ and to get deeper insight into chemical bonding in the metallic phases of FeH₃ and FeH₄ described above, we use Brown's bond valence method [61], the main formula of which is:

$$n = \exp(\frac{R_0 - R}{b}) \quad , \tag{1}$$

where $n$ is the bond valence, $R$ is the bond length, b a constant (usually taken as 0.37 Å), and $R_0$ a bond-specific constant (which, however, depends on pressure). The sum of bond valences then defines the valence of the atom. To apply eq. (1), we first calibrate $R_0$ values on a structure where the valences are known – FeH. We obtain $R_0 = 1.32$ (1.27) Å for Fe-Fe, 0.92 (0.89) Å for Fe-H and 0.51 (0.50) Å for H-H bonds at 300 GPa (400 GPa). The values for Fe-Fe and Fe-H bonds are obtained directly, while the one for H-H is deduced from them and is therefore less accurate.

Applying eq. (1) to FeH₃, we obtain the sum of bond valences on Fe equal to 2.6 at 300 GPa and 2.7 at 400 GPa – in reasonable agreement with the expected valence of 3. Turning to FeH₄, at 400 GPa again obtain the sum of bond valences equal to 2.7 – i.e. again trivalent iron, a prediction that can be verified experimentally using Mössbauer spectroscopy. Bond valence calculations also explain how FeH₄ can have trivalent (rather than tetravalent) iron – extensive H-H bonding satisfies part of their valence needs and decreases the need for additional electrons from Fe atoms. Very crudely, half of hydrogen's valence is satisfied by H-H bonds, and to satisfy valence of four hydrogen atoms only 2 (instead of 4) electrons from Fe are needed. Each Fe atom uses two valence electrons for Fe-H bonding, and 1 additional electron is used for Fe-Fe bonding – making the total valence of Fe atoms equal to 3.

Stable hydrides that contain more H than prescribed by naive application of chemical valence have been predicted, e.g. for the Li-H system under pressure [62], and include such exotic compounds as LiH₂, LiH₆ and LiH₈. Just like in FeH₄, those compounds also contain H-H bonds. Even in hydrides of "normal" stoichiometries, such as GeH₄ [63] and SnH₄ [64] high pressure promotes the formation of H-H bonds that should be accompanied by a decrease of the metal valence (we showed that in the high-pressure *Cmcm* phase of SnH₄ tin is counterintuitively divalent, rather than tetravalent [50]).

## 5. How much carbon and hydrogen is needed to explain the density of the inner core?

If C and H are to be considered as potential major light elements in the core, several conditions have to be met – (i) the amount of light element needed to explain the observed core density at the expected core temperatures (5000-6000 K [65]) should not be unacceptably large (roughly, < 20mol.%), (ii) this amount should not display large and non-monotonic variations with depth, (iii) the resulting mean atomic mass $\overline{M}$ should be reasonably close to the one determined using the Birch's law, 49 [5]. As we show below, carbon satisfies all these necessary conditions, which means that it is a good candidate to be a major alloying element in the inner core. Hydrogen satisfies conditions (i) and (ii), but not (iii).

Our procedure is as follows. Since theoretical absolute densities suffer from small, but non-negligible systematic errors of theory and accurate experiment-based *P-V-T* equation of state is known for pure hcp-iron [66], we based our estimates on the known density of pure iron at relevant pressures and temperatures and treated the compositional effect on the density as linear, and determined it from the theoretical density differences between hcp-Fe and stable iron carbide (Fe$_2$C) or hydride (FeH) at relevant pressures and *T* = 0 K. Parameters of the relevant theoretical equations of state are given in Table 2 and are close to the previous theoretical values [20, 29, 70].

**Table 2. Theoretical third-order Birch-Murnaghan equations of state of the non-magnetic high-pressure phases in the Fe-C and Fe-H systems. Theoretical data refer to the 0 K isotherms without zero-point motion. For diamond, experimental data [67] are shown in parentheses. For comparison of theoretical and experimental equations of state, see also Fig. 1.**

| $V_O$, Å$^3$/atom | $Ko$, GPa | $Ko'$ |
|---|---|---|
| | hcp-Fe | |
| 10.15 | 305.7 | 4.3 |
| | Fe$_3$C-cementite | |
| 8.88 | 326.1 | 4.31 |
| | $C2/m$-Fe$_3$C | |
| 8.97 | 283.2 | 4.56 |
| | $I\bar{4}$-Fe$_3$C | |
| 8.78 | 333.6 | 4.34 |
| | Fe$_7$C$_3$ | |
| 8.68 | 317.6 | 4.37 |
| | Fe$_2$C | |
| 8.44 | 333.9 | 4.23 |
| | Diamond | |
| 5.71 (5.68) | 431.8 (446) | 3.62 (3) |
| | fcc-FeH | |
| 6.11 | 270.8 | 4.25 |

We determine the molar concentration of the light element needed to explain the density of the inner core by matching the observed density of the inner core to the density of the mixture of hcp-Fe and the stable carbide/hydride relevant pressures and temperatures. This corresponds to the concentration needed if carbon or hydrogen were the only light alloying element and, since several alloying elements are likely to the present in different concentrations, gives the upper bound for the concentration of each element. For instance, for the case of carbon, considering the Fe-Fe$_2$C mixture, we could determine the molar concentration of carbon that matches the two densities:

$$\rho_{IC} = \rho_{Fe}^T + \frac{\partial \rho}{\partial x} \cdot x \Rightarrow \rho_{IC} - \rho_{Fe}^T = \frac{\rho_{Fe_2C}^{0K} - \rho_{Fe}^{0K}}{0.33} \cdot x, \qquad (2)$$

where $\rho_{IC}$ is the PREM [68] density of the inner core at each depth, $\rho^T{}_{Fe}$ is the density of pure iron at given temperature [66], $\rho^{0K}{}_{Fe}$ and $\rho^{0K}{}_{Fe2C}$ are the computed zero-Kelvin densities of Fe and Fe$_2$C. The number 0.33 in Eq. (2) indicates the molar fraction of carbon in Fe$_2$C. The essence of eq. (2) is compute the *PVT*-equation of state of a Fe-C(H) alloy from the well-constrained *PVT*-equation of state of pure iron [66], supplementing it with the theoretically computed (at zero Kelvin) compositional derivatives of the density. This is a trick to solve a well-recognized problem: density-functional calculations have systematic errors equivalent to the shift of the absolute equation of state by several GPa [69], which makes direct comparison of theoretical equations of state with seismic data dangerous. However, volume (or density) differences (such as the effects of the composition or temperature on the density) are usually very accurate due to compensation of errors, and thus Eq. (2) gives reliable estimates.

The resulting concentrations are (estimated along to isotherms, 5000 K and 6000 K) are very reasonable, 11-15 mol.% C (2.6-3.7 weight % C), and do not show large variations throughout the inner core. This matches the concentration of carbon in CI carbonaceous chondrites. Furthermore, the resulting $\overline{M}$ is in the range 49.3-51.0, which is very close to the desired value of 49 [5]. All this indicates that a significant amount of carbon can exist in the Earth's inner core. Note that these estimates for carbon are not very sensitive to the choice of the reference carbide: Fig 7 shows that if we take $Fe_3C$ or $Fe_7C_3$ instead of $Fe_2C$. Our estimates are compatible with the latest estimates based on the equation of state of $Fe_7C_3 - 3.2$ wt.% [28], 1.5 wt.% [29], 3.7 wt.% (Prakapenka, pers.comm.).

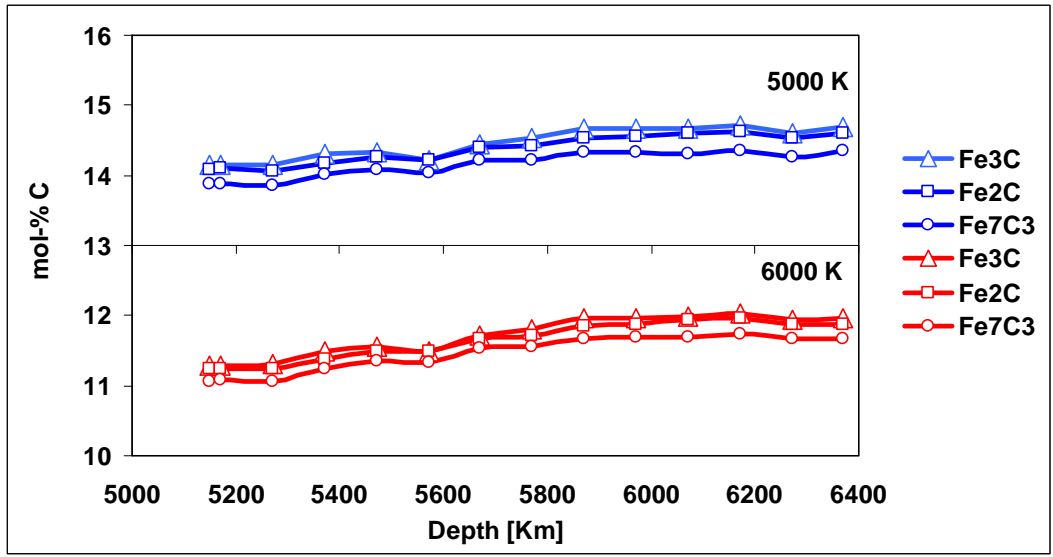

**Fig. 7. Matching molar concentration of carbon as a function of depth in the inner core.** Results based on $Fe_3C$, $Fe_7C_3$ and $Fe_2C$ reference carbon-bearing phases are shown along the 5000 K and 6000 K isotherms.

Similarly made estimates for the hydrogen matching concentrations are shown in Fig. 8. The matching concentrations vary between 17-22 mol.% (0.4-0.5 wt.%) if we use the equation of state of FeH (or 24–32 mol.% if we used $Fe_4H$). This is much higher than the recent estimate 0.08-0.16 wt.% by Narygina et al. [46], but is compatible with the 0.12-0.48 wt.% proposed by Hirao et al. [37]. Yet, hydrogen concentrations of 17-22 mol.% appear too large to be realistic, especially in view of the average atomic mass $\overline{M}$ in the range 43.8-46.5, which is too low compared to $\overline{M}$ =49 inferred from the Birch's law. We cannot rule out the presence of considerable amounts of hydrogen in the inner core, but it certainly cannot be the dominant light element.

At first sight, it might appear surprising that hydrogen, being much lighter than carbon, is required in greater concentrations than carbon to create the observed density deficit. This becomes less surprising if one considers crystal chemistry of iron carbides and hydrides at high pressures. Carbon is much larger and significantly affects the crystal structure of the alloy, where it occupies rather large sites with 8- or 9-fold coordination. All this has significant effects for the density - because insertion of carbon destroys close packing of the Fe atoms and significantly increases the unit cell volume. On the other hand, much smaller hydrogen sits comfortably in the octahedral voids of the close packing of iron atoms (in our structure searches, we also saw a metastable structures hosting hydrogen in even smaller voids). Incorporation of hydrogen does not destroy the close packing of the Fe atoms and has only a minor effect of the unit cell volume. Its effect on the density is about two times smaller than that of carbon incorporation.

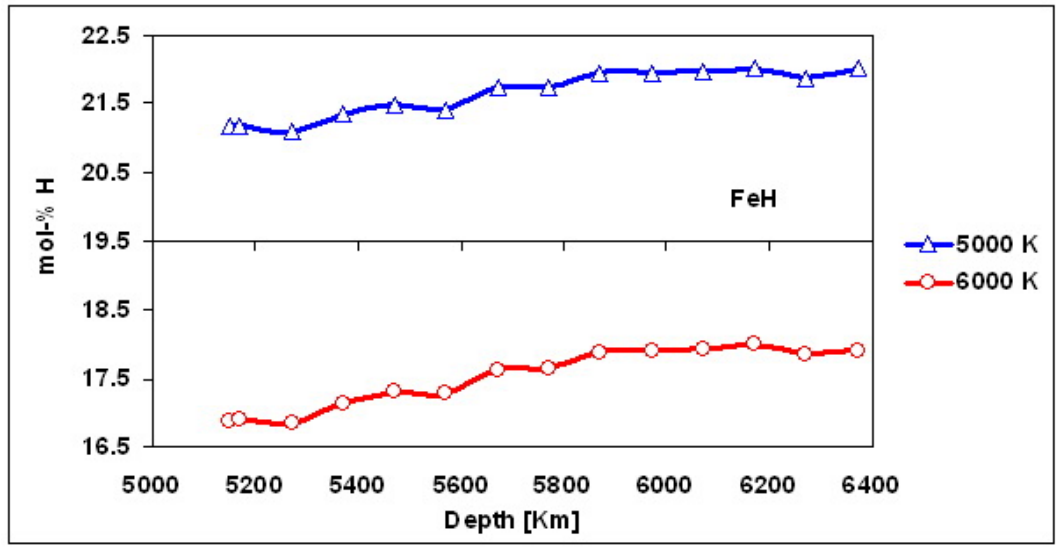

**Fig. 8. Matching molar concentration of hydrogen as a function of depth in the inner core.** Results are shown along the 5000 K and 6000 K isotherms.

## 6. Conclusions

The problem of the chemical composition of the Earth's inner core has occupied the minds of scientists for decades, and in recent years there has been resurgence of interest in carbon and hydrogen as potential light alloying elements in the iron-dominated core. The results obtained by different researchers differed greatly, because of the assumptions and approximations made. Using state-of-the-art *ab initio* simulation techniques, including evolutionary crystal structure prediction and density functional theory, new insight has been obtained for the Fe-C and Fe-H systems at pressures of the Earth's inner core. Evolutionary crystal structure prediction method USPEX was shown to be more reliable than the random sampling method in searching for ground state structures. New hitherto unsuspected ground states have been found using USPEX in both systems. Chemically (though not mineralogically) important and surprising tendency is the destabilization of $FeH_2$ and stabilization of $FeH_4$ under pressure; stability of even higher iron hydrides has not been investigated here and certainly deserves a separate study. Crystal structures of iron carbides and hydrides at inner core pressures show a striking difference - while hydrogen atoms sit in the interstices of close packed iron structures and have minor effect on the density, carbon atoms occupy much larger sites in complex non-close packed structures and as a consequence carbon has a much stronger effect on the density. Therefore, to match the observed density of the inner core, one needs a much greater and unrealistic concentration of hydrogen. At the same time, carbon (at least from the point of view of density of the inner core) cannot be ruled out as the dominant light element in the inner core (with the concentration in the range of 11-15 mol.%). Coupled with geochemical and meteoritic evidence, the presence of significant amounts of carbon in the Earth's core seems plausible, logical and perhaps even unavoidable.

Then, if carbon and hydrogen cannot simultaneously be present in the Earth's core, as recently claimed [46], then presence of hydrogen could be ruled out with great probability. But before that, conclusions of [46] need to be re-assessed. Further constraints can be provided by the density and compressional wave velocity of the liquid outer core, seismic wave velocities in the inner core, and chemical equilibria at the inner-outer core boundary [71]. Systematic analysis of the effect and possible presence of other elements (S, Si, O), based on crystal structure prediction, is urgently needed. The results on the Fe-Si system have already been published [7172] and studies of the Fe-O and Fe-S systems are underway.

## Acknowledgments

This work is supported by the U.S. National Science Foundation (grant EAR-1114313) and DARPA (grant N66001-10-1-4037). Calculations were performed at Moscow State University (Skif MSU supercomputer), at the Joint Supercomputer Center of the Russian Academy of Sciences, and on the CFN cluster (Brookhaven National Laboratory), which is supported by the U.S. Department of Energy, Office of Basic Energy Sciences, under contract No. DE-AC02-98CH10086.